\documentstyle[epsfig,12pt,amsmath,amssymb]{article}
\def\mol{D^0\bar D^{*0}}
\newcommand{\bbr}{{\it B{\footnotesize A}B{\footnotesize AR}}}
\def\lclc        {{\Lambda_c^+\overline{\Lambda}_c^-}}
\def\psii       {{\Psi(2S)}}
\def\y       {${Y(4008)}$}
\def\yy       {${Y(4260)}$}
\def\yyy       {${Y(4325/4360)}$}
\def\yyyy       {${Y(4660)}$}
\def\x       {${X(4630)}$}

\def\bes{{\it BES III}}
\def\bepcii{{\it BEPC II}}

\newcommand{\cc}{$c\bar{c}$}

\newcommand{\mevc}{$\, {\mathrm{MeV}/c^2}$}
\newcommand{\gev}{$\, {\mathrm{GeV}}$}
\newcommand{\mev}{$\, {\mathrm{MeV}}$}

\newcommand{\jp} {$J/\psi$}
\newcommand{\pp} {$\psi^\prime$}

\newcommand{\eejpet} {$e^+ e^- \! \to \! J/\psi \, \eta_c$}
\newcommand{\eejpgg} {$e^+ e^- \! \to \! J/\psi \, gg$}
\newcommand{\eejpg}  {$e^+ e^- \! \to \! J/\psi \, g$}
\newcommand{\eejpcc} {$e^+ e^- \! \to \! J/\psi \, c\bar{c}$}

\newcommand{\eejpxn} {$e^+ e^- \! \to \! J/\psi \, X(3940)$}

\newcommand{\jpncc} {$J/\psi \, X_{\text{non-}c\bar{c}}$}

\newcommand{\jpcc} {$J/\psi \, c\bar{c}$}

\newcommand{\dd}   {$D \overline{D}$}
\newcommand{\dds}  {$D^* \overline{D}$}
\newcommand{\dsds} {$D^* \overline{D}{}^*$}

\newcommand{\eedd}   {$e^+ e^- \! \to \! J/\psi \, D \overline{D}$}

\newcommand{\eedsds} {$e^+ e^- \! \to \! J/\psi \, D^* \overline{D}{}^*$}

\newcommand{\ee}{e^{+} e^{-}}

\newcommand{\psip}{\psi '}
\newcommand{\jpsi}{J/\psi}

\newcommand{\pipi}{\pi^{+}\pi^{-}}

\newcommand{\rt}{\rightarrow}

\begin{document}

\thispagestyle{empty}

$\phantom{.}$

\hfill
\includegraphics*[width=40mm]{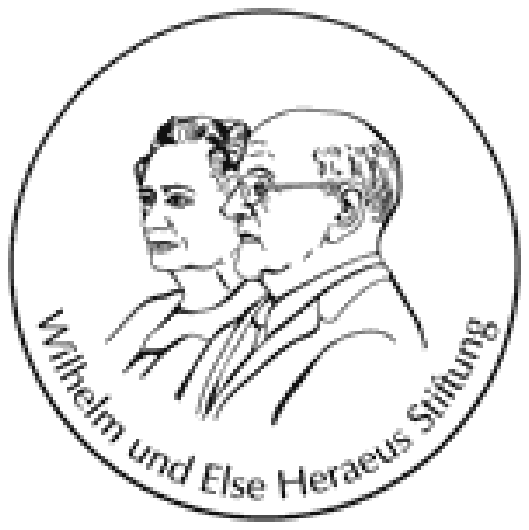}

\begin{center}
{\Large {\bf 447$^{\rm th}$ Wilhelm and Else Heraeus Seminar:} \\
\vspace{0.75cm}
{\huge {\bf  Charmed Exotics}}}

\vspace{1cm}

{\large August 10--12, 2009, Bad Honnef, Germany}

\vspace{1cm}


\vspace{1cm}

{\it Editors:} {\bf G.~Bali}~(Regensburg),
    {\bf A.~Denig}~(Mainz),
{\bf S.I.~Eidelman}~(Novosibirsk),
{\bf C.~Hanhart}~(J\"ulich),
{\bf S.~Krewald}~(J\"ulich),
  {\bf U.-G.~Mei\ss ner}~(Bonn/J\"ulich),
{\bf A.~Sibirtsev}~(Bonn/JLab), and
    {\bf U.~Wiedner}~(Bochum) 

\vspace{2.5cm}

ABSTRACT

\end{center}

\vspace{0.3cm}

\noindent
These are the mini-proceedings 
of the {\rm CHARMEX} workshop. 
The meeting focused on recent developments in charm
spectroscopy, especially on the possible role of the states
that do not fit into the quark model classification ---
the so--called exotic states.
The goal of this
write-up is to provide the community with a short summary
of the individual talks as well as a comprehensive, up--to--date list
of relevant references.

\newpage

{$\phantom{=}$}

\vspace{0.5cm}

\tableofcontents

\newpage

\section{Introduction}

\subsection{Scope of the workshop}
\addtocontents{toc}{\hspace{2cm}{\sl C.~Hanhart}\par}

\vspace{5mm}

C.~Hanhart

\vspace{5mm}

\noindent
Institut f\"ur Kernphysik (Theorie) and J\"ulich Center for Hadron
Physics, Forschungszentrum J\"ulich

\vspace{5mm}
Until the turn of the millennium there was a strong belief that
hadrons containing the charm quark and among those especially the
charmonia (states with a charm and an anticharm quark) can be largely
understood on the basis of the non--relativistic quark
model. Corrections were believed to be calculable within Heavy Quark
Effective Field Theory (HQEFT). However, especially since BaBar and
BELLE started to enter the field of charm spectroscopy, a large number
of new states were discovered that because of their mass and/or their
properties do not at all match the described picture.  How can we
understand their nature in terms of QCD? Can they be included in the
quark model after some refinements or are they of completely different
origin --- options suggested in the literature range from glueballs
over hybrids to molecules?  How can one distinguish among those
different scenarios?  What are the proper theoretical tools for the
analysis of these states?

\medskip\noindent
To address these questions is especially important now, when the
components of the FAIR projects are in their final phase of
planning. Charm spectroscopy will be studied in the PANDA
experiment at the High Energy Storage Ring. Insights on relevant
observables and requests on the resolution or particle identification
that might emerge from nowadays discussions can still influence some
aspects of the detector.  In this context it is very important to
identify and further refine, on the basis of what can be expected in
the near future from Belle, BaBar, BES-III, CLEO-c, D0, and CDF, the minimal
requirements for the PANDA detector to make sure that the experiment
will indeed improve our understanding of QCD.

\medskip\noindent
The webpage of the conference, which contains all talks, can be found under
\begin{center}
www.fz-juelich.de/ikp/charmex
\end{center}

\vspace{0.5cm}

\noindent
The meeting would not have been possible without the support
by the Wilhelm and Else Heraeus foundation. All participants
especially appreciated the efficient and unbureaucratic
style of the foundation. The  Wilhelm and Else Heraeus foundation
is the most important private foundation to support natural sciences.
For more information see
http://www.we-heraeus-stiftung.de/~. We would also like to thank
the staff of the Physikzentrum, in particular V.~Gomer, for the 
efficient organization.

\newpage

\section{Short summary of the talks}

\subsection{Exotics at Belle and BaBar}
\addtocontents{toc}{\hspace{2cm}{\sl S.L.~ Olsen}\par}

\vspace{5mm}

\noindent
Stephen L. Olsen

\vspace{5mm}

\noindent
   Seoul National University
           Seoul KOREA

\vspace{5mm}

A review of some of the recent experimental developments concerning 
the $X$, $Y$ and $Z$ charmoniumlike mesons states is presented.  
New mass measurements from Belle and CDF place the mass of the
$X(3872)$ at $0.35\pm0.41$~MeV below the $m_{D^0}+m_{D^{*0}}$
threshold. No strong evidence if seen for $B\rt K^*(890)X(3872)$
in contrast to all known charmonium states for which strong signals for
$B\rt K^*(890)+$charmonium are seen.
 Belle reports a new, near-threshold $\omega$\jp
mass peak in $\gamma\gamma\rt\omega$\jp decays, with mass and width
consistent with BaBar's recent measured values for the $Y(3940)$
from $B\rt K\omega$\jp decays.
A Belle search for $Y(3940)\rt D^*\bar{D}$ resulted in an upper
limit that contradicts earlier measurements for the $X(3940)$
(a $D^*\bar{D}$ peak seen in $\ee\rt$\jp$D^*\bar{D}$ annihilation),
thereby establishing the $Y(3940)$ and $X(3940)$ as distinct
states.  No evidence is seen for the $1^{--}$ $Y$ states in any
open charmed decay channels, including the $D^{**}\bar{D}$
channels favored by $c\bar{c}$-gluon hybrid models.
A Belle reanalysis of $B\rt K\pi^+\psip$ decays
using Dalitz techniques confirms their earlier claims for
a charged $Z(4430)^+\rt \pi^+\psip$ resonance.

\noindent


\newpage

\subsection{Exotic Charmonia}
\addtocontents{toc}{\hspace{2cm}{\sl E.~ Braaten}\par}

\vspace{5mm}

\noindent
Eric Braaten

\vspace{5mm}

\noindent
Department of Physics, The Ohio State University

\vspace{5mm}

In this talk, I discussed the $c \bar c$ mesons above the 
$D \bar D$ threshold that have been discovered in recent years.
I summarized the various proposals for identifying some of these 
states as exotic $c \bar c$ mesons.
I then focused on a specific state that is definitely an 
exotic $c \bar c$ meson:  the $X(3872)$.
I explained how existing data implies unambiguously
that this state is a loosely-bound charm-meson molecule.
I also discussed various misconceptions that have preventing this
identification from being universally accepted in the 
particle physics community.

In the references below, I list some review articles on
$c \bar c$ mesons above the $D \bar D$ threshold
\cite{Swanson:2006st,Eichten:2007qx,Voloshin:2007dx,Godfrey:2008nc,%
Braaten:2008nv}.
I also list my papers on the $X(3872)$ in which I have been 
developing the case for this state as a   
loosely-bound charm-meson molecule
\cite{Braaten:2003he,Braaten:2004fk,Braaten:2004jg,Braaten:2004ai,%
Braaten:2005jj,Braaten:2005ai,Braaten:2006sy,Braaten:2007dw,%
Braaten:2007ft,Braaten:2007sh,Stapleton:2009ey}.

\newpage

\subsection{Selected Topics in BES and Belle experiments}
\addtocontents{toc}{\hspace{2cm}{\sl C.Z. Yuan}\par}

\vspace{5mm}

\noindent
Chang-Zheng Yuan

\vspace{5mm}

\noindent
Institute of High Energy Physics, Chinese Academy of
Sciences, Beijing 100049, China

\vspace{5mm}

The following topics were discussed: {\bf (1)} the $Y(2175)$
signals observed in $J/\psi\to \phi f_0(980) \eta$ at
BES~\cite{besy2175} and in $e^+e^-\to
\phi\pi^+\pi^-$~\cite{belley2175} at Belle experiments; {\bf (2)}
the anomalous structure in the total cross section around
3.77~GeV~\cite{bes2pspp} in the $e^+e^-$ annihilation observed at
BESII experiment; {\bf (3)} the $X(3915)$ particle observed in
two-photon process $\gamma\gamma\to \omega
J/\psi$~\cite{bellex3915}: Belle observed a resonance-like
enhancement in $\gamma\gamma\to \omega J/\psi$ process with a
statistical significance of $7.7\sigma$ composed of $55\pm 14$
events in the peak region. The mass and width are measured to be
$M = 3914\pm 3\pm 2$~MeV/$c^2$, and $\Gamma = 23\pm
10^{+3}_{-8}$~MeV, respectively. It is noted that these values are
close to those of the $Y(3940)$ from the BaBar
measurent~\cite{babary3940}. The total width from the same kind of
the Belle result seems to be a little larger~\cite{belley3940}.
This state may be identical to either $Y(3940)$ or
$Z(3930)$~\cite{bellez3930} but not very conclusive; {\bf (4)} the
evidence for a new resonance $X(4350)$ in two-photon process
$\gamma\gamma\to \phi J/\psi$~\cite{bellex4350} at Belle
experiment: Evidence is reported for a narrow structure at
$4.35~\hbox{GeV}/c^2$ in the $\phi J/\psi$ mass spectrum in
two-photon process $\gamma^* \gamma^* \to \phi J/\psi$. The
analysis is based on a data sample of 825~fb$^{-1}$ collected on
and off the $\Upsilon(nS)~(n=1,3,4,5)$ resonances with the Belle
detector. A signal of $8.8^{+4.2}_{-3.2}$ events, with statistical
significance of 3.9 standard deviations, is observed. The mass and
natural width of the structure (named as $X(4350)$) are measured
to be $4350.6^{+4.6}_{-5.1}(\rm{stat})\pm
0.7(\rm{syst})~\hbox{MeV}/c^2$ and
$13.3^{+17.9}_{-9.1}(\rm{stat})\pm 4.1(\rm{syst})~\hbox{MeV}/c^2$,
respectively. The products of its two-photon decay width and
branching fraction to $\phi J/\psi$ is measured to be
$\Gamma_{\gamma \gamma}(X(4350)) B(X(4350)\to\phi
J/\psi)=6.4^{+3.1}_{-2.3}\pm 1.1~\hbox{eV}$ for $J^P=0^+$, or
$1.5^{+0.7}_{-0.5}\pm 0.3~\hbox{eV}$ for $J^P=2^+$. No $Y(4143)$
signal~\cite{CDF} is observed, and $\Gamma_{\gamma
\gamma}(Y(4143)) B(Y(4143)\to\phi J/\psi)<39~\hbox{eV}$ for
$J^P=0^+$ or $<5.7~\hbox{eV}$ for $J^P=2^+$ is determined at the
90\% C.L.; and {\bf (5)} the status of the BESIII experiment at
the BEPCII: BESIII started taking data for physics study since
spring 2009. Up to now, 107~M $\psi(2S)$ events and about 200~M
$J/\psi$ events have been accumulated. There is a rich physics
program at BESIII experiment~\cite{bes3}.

\newpage

\subsection{Strong Decays on the Lattice}
\addtocontents{toc}{\hspace{2cm}{\sl C.~ McNeile}\par}

\vspace{5mm}

\noindent
Craig McNeile

\vspace{5mm}

\noindent
University of Wuppertal

\vspace{5mm}

I started the talk with a brief discussion
of the status of lattice QCD~\cite{Jansen:2008vs}.
I reported on precision results from lattice QCD calculations,
such as the mass of the 
charm quark ($m_c(m_c)$=1.268(9)GeV)~\cite{Allison:2008xk}, 
and the strong coupling 
($\alpha_{\overline{MS}}(M_Z,n_f=5)$= 0.1183(8))~\cite{Allison:2008xk}.

I reviewed lattice QCD calculations of: 
the $\rho$ coupling to 2$\pi$~\cite{McNeile:2002fh}, 
$b_1 \rightarrow  \pi \omega$~\cite{McNeile:2006bz}, 
light $1^{-+} \rightarrow \pi b_1$~\cite{McNeile:2006bz},
and
light $1^{-+} \rightarrow \pi f_1$~\cite{McNeile:2006bz}.
Other lattice calculations of strong decays were
included in the summary tables~\cite{McNeile:2002az}.
I reviewed lattice QCD calculations of the $g_{B B^\star\pi}$
coupling~\cite{Abada:2002xe}.
I reported on a lattice QCD of the strong decay of 
the $B^{\star\star}$
meson~\cite{McNeile:2004rf}.

I ended the talk reviewing recent lattice QCD
calculations that use L\"{u}scher's method~\cite{Luscher:1991cf}
(and variants of) to study the 
properties of the $\rho$ 
and $\Delta$~\cite{Aoki:2007rd}.

\newpage

\subsection{Excited Charmonium and radiative transitions from
lattice QCD}
\addtocontents{toc}{\hspace{2cm}{\sl J. ~Dudek}\par}

\vspace{5mm}

\noindent
Jozef J. Dudek

\vspace{5mm}

\noindent
Jefferson Laboratory, 12000 Jefferson Avenue,  Newport News, VA 23606, USA\\
Department of Physics, Old Dominion University, Norfolk, VA 23529, USA\\

\vspace{5mm}

Using point-all technology on quenched lattices, the excited spectrum of charmonium \cite{spec} and radiative transition amplitudes between excited states \cite{rad1, rad2} are computed. That this is possible follows from application of a large basis of composite QCD interpolating fields. Highlights include the first extraction from QCD of a radiative transition featuring an exotic quantum numbered state. Phenomenological implications of these results are presented \cite{ermal}.

Preliminary results in the light meson spectrum are also presented. These follow from the use of distillation technology \cite{distillation} on dynamical anisotropic lattices \cite{aniso}. The extracted spectrum at the strange quark mass shows many of the systematics of the quark model with the addition of a number of exotic and non-exotic states which appear to have some properties expected of hybrid mesons \cite{light_spec}.

\newpage

\subsection{QCD Exotics at BNL and JLab}
\addtocontents{toc}{\hspace{2cm}{\sl C.~ Meyer}\par}

\vspace{5mm}

\noindent
Curtis A. Meyer

\vspace{5mm}

\noindent
   Carnegie Mellon University

\vspace{5mm}

Searches have been carried out at Brookhaven National Lab looking for light-quark mesons
with non-quark-antiquark ($q\bar{q}$) quantum numbers. Lattice 
calculations~\cite{lacock-97,bernard-97,bernard-99,lacock-98,zhong-03,bernard-04}
predict that several nonets of these mesons should exist, and phenomenology gives some
guidance on how they should decay~\cite{page-99}. The first such state reported is an
isospin 1 $J^{PC}=1^{-+}$ state ($\pi_{1}(1400)$) with a mass near $1.4$~GeV and a width 
of around $0.3$~GeV~\cite{Thompson:1997bs,Chung:1999we}. This state was also
observed in $\bar{p}n$ annihilation at rest by Crystal Barrel~\cite{Abele:1998gn,Abele:1999tf}.
An alternate analysis of more E852 data on both $\pi^{-}p\rightarrow n \eta\pi^{0}$ and 
$p\eta\pi^{-}$ found no evidence for the $\pi_{1}$~\cite{Szczepaniak:2003vg}. However,
a later E852 analysis on $\pi^{-}p\rightarrow n \eta\pi^{0}$ confirmed their earlier 
result~\cite{Adams:2006sa}. E852 also found evidence for a second $\pi_{1}$ state
in several final states: $\pi^{-}p\rightarrow p\pi^{+}\pi^{-}\pi^{-}$\cite{Adams:1998ff,Chung:2002pu},
$\pi^{-}p\rightarrow p \eta^{\prime}\pi^{-}$\cite{Ivanov:2001rv}, 
$\pi^{-}p\rightarrow p \omega \pi^{0}\pi^{-}$\cite{Lu:2004yn} and
$\pi^{-}p\rightarrow p f_{1}(1285)\pi^{-}$\cite{Kuhn:2004en}. This new state appears to 
have a mass of about $1.6$~GeV and a width $\sim 0.3$~GeV. However, the observed
production mechanism is not consistent over all the reported observations. Interestingly,
an alternate analysis of a more extensive E852 data set seems to indicate that in the 
$3\pi$ final state, the observed exotic wave is actually leakage from well-known decays 
of the $\pi_{2}(1670)$~\cite{Dzierba:2005jg}. The $\pi_{1}(1600)$ was also looked for 
in photo production by CLAS in $\gamma p \rightarrow n\pi^{+}\pi^{+}\pi^{-}$~\cite{:2008bea},
but no evidence was found. Finally, in the $f_{1}\pi$ and $b_{1}\pi$ final
states, E852 reports weak evidence for a 3'rd exotic state, the $\pi_{1}(2000)$. 

In the future, we anticipate exciting results from the GlueX~\cite{gluex} which will
use $8.4-9$~GeV linearly polarized photons incident on protons to photoproduce 
hydrid mesons. With GlueX, we hope to map out the three exotic meson nonets 
expected from theory. Construction of the energy doubling upgrade of Jefferson Lab
as well as GlueX started late in 2008 and first beam is anticipated in GlueX in 2014.

\newpage

\subsection{Recent Bottonomium results from BaBar}
\addtocontents{toc}{\hspace{2cm}{\sl V.~ Ziegler}\par}

\vspace{5mm}

\noindent
Veronique Ziegler

\vspace{5mm}

\noindent
SLAC National Accelerator Laboratory

\vspace{5mm}

A search for the bottomonium ground state $\eta_b(1S)$
in the photon energy spectrum was performed using a sample of $(109 \pm 1)$ million of $\Upsilon(3S)$
recorded at the $\Upsilon(3S)$
energy with the BaBar detector at the PEP-II $B$ factory at SLAC~\cite{etabfrom3S}.
We observe a peak in the photon energy spectrum at $E_\gamma = 921.2 ^{+2.1}_{-2.8} {\rm (stat)}\pm 2.4{\rm (syst)}$ MeV
with a significance of 10 standard deviations. We interpret the observed peak as being due to
monochromatic photons from the radiative transition $\Upsilon(3S) \to \gamma \, \eta_b(1S)$.
This photon energy corresponds to an $\eta_b(1S)$ mass of $9388.9 ^{+3.1}_{-2.3} {\rm (stat)} \pm 2.7{\rm (syst)}$ MeV/$c^2$.
The hyperfine $\Upsilon(1S)$-$\eta_b(1S)$ mass splitting is $71.4 ^{+2.3}_{-3.1} {\rm (stat)} \pm 2.7{\rm (syst)}$ MeV/$c^2$.
The branching fraction for this radiative $\Upsilon(3S)$ decay is estimated to be
$(4.8 \pm 0.5{\rm (stat)}  \pm 1.2 {\rm (syst)}) \times 10^{-4}$.
A similar strategy was utilized to  search for the $\eta_b(1S)$ meson
in the radiative decay of the $\Upsilon(2S)$ resonance using a sample of 91.6 million $\Upsilon(2S)$ events~\cite{etabfrom2S}.
A peak was observed in the photon energy spectrum at $E_\gamma = 609.3 ^{+4.6}_{-4.5} {\rm (stat)}\pm 1.9{\rm (syst)}$ MeV,
corresponding to an $\eta_b(1S)$ mass of $9394.2 ^{+4.8}_{-4.9} {\rm (stat)} \pm 2.0{\rm (syst)}$ MeV/$c^2$.
The branching fraction for the decay $\Upsilon(2S)\rightarrow\gamma\eta_b(1S)$ is determined to be
$(3.9 \pm 1.1 {\rm (stat)}  ^{+1.1}_{-0.9} {\rm (syst)}) \times 10^{-4}$.
We find the ratio of branching fractions ${\cal B}(\Upsilon(2S)\rightarrow\gamma\eta_b(1S))/{\cal B}(\Upsilon(3S)\rightarrow\gamma\eta_b(1S)) = 0.82 \pm 0.24 {\rm (stat)}$$ ^{+0.24}_{-0.20}{\rm (syst)}$.

Between March 28 and  April 7, 2008
the PEP-II $e^+ e^-$ delivered colliding beams at
a center-of-mass energy ($\sqrt{s}$) in the range of 10.54 to 11.20 GeV. First, an
energy scan over the whole range in 5 MeV steps, collecting approximately
25 pb$^{-1}$ per step for a total of about 3.3 fb$^{-1}$, was performed~\cite{Scan}.
It was then followed by a 600 pb$^{-1}$ scan in the range of $\sqrt{s}$=10.96 to 11.10
GeV, in 8 steps with non-regular energy spacing, performed in order to
investigate the $\Upsilon(6S)$ region. This data set outclasses the previous
scans~\cite{Y_CLEO,Y_CUSB} by a factor $> 30$ in the luminosity and $\sim 4$ in
the size of the energy steps.
For each step in $\sqrt{s}$,  $e^+e^-\to b\bar{b}$ cross section measurements were obtained.
A total relative error of about 5\% is reached in more than 300
center-of-mass energy steps, separated by  about 5 MeV.
These measurements can be used to derive
precise information on the parameters of the $\Upsilon(5S)$ and $\Upsilon(6S)$
and have the potential to yield information on the bottomonium
spectrum and possible exotic extensions.

\newpage

\subsection{Unquenching, Requenching, and Renormalising the Quark Model}
\addtocontents{toc}{\hspace{2cm}{\sl E.~ Swanson}\par}

\vspace{5mm}

\noindent
Eric S. Swanson

\vspace{5mm}

\noindent
Department of Physics and Astronomy, University of Pittsburgh

\vspace{5mm}

\noindent
The issue of incorporating virtual quark effects into the constituent quark model
has become more germane with the recent charmonium discoveries in the continuum region.
Nevertheless, the importance of this effect has been recognised for a long time\cite{OY}.
This is a difficult problem that is probably not amenable to an effective field theory
approach, and relies on guessing nonperturbative gluodynamics that drives quark pair
production in the soft regime. One such guess is the $^3P_0$ model\cite{3p0}, but others
based on one gluon exchange\cite{3S1}, or a relativistic kernel\cite{ccc} exist.

Incorporating unquenching effects is not technically easy\cite{ess}, but many computations have been
made\cite{unquenching}. A recent advance is the establishment of theorems that guarantee
that unquenching induced mass splittings will be identical for meson in degenerate SU(6) 
multiplets\cite{bs}. This lends support to the notion that the constituent quark model
should be stable with respect to spin splitting effects, however direct computations indicate large residual shifts that cannot be absorbed in the model parameters. 

An alternative approach is to incorporate unquenching effects in an  effective interaction
by implementing a similarity transformation in powers of the inverse quark mass\cite{ess2}.
This yields a spin-independent quark-quark effective interaction and a spin-orbit quark-antiquark
interaction, with interesting implications for baryon spectroscopy.

Finally, unquenching quark models necessitates implementing (finite) renormalisation.
This applies to typical quark model parameters, which now must be considered cut-off 
dependent, and to the quark charge.

\vspace{5mm}

\newpage

\subsection{Nature of $X(3872)$ from data}
\addtocontents{toc}{\hspace{2cm}{\sl A. V. ~Nefediev}\par}

\vspace{5mm}

\noindent
Alexei V.  Nefediev

\vspace{5mm}

\noindent
Institute of Theoretical and Experimental Physics,\\
117218, B.Cheremushkinskaya 25, Moscow, Russia

\vspace{5mm}

The nature of the state $X(3872)$ is discussed as based on the recent
experimental data \cite{xnew}. In particular, the data on the $D\bar{D}^*$
and $\pi^+\pi^- J/\psi$ channels, as provided by the Belle
Collaboration \cite{Belle3875v2,Belle2} and by the BaBar Collaboration
\cite{BaBar2,Babarddpi}, are analysed using the method suggested in
\cite{recon} and based on the Flatt{\'e} parametrisation of the $D\bar{D}^*$
amplitude. The method is generalised to include into consideration extra decay
channels for the $X$, in particular, radiative decay channels \cite{BaBarrad}.
In
addition, the effect of the finite $D^*$ width and the interference in the
$D\bar{D}^*$ system is discussed. The conclusion is made that the
$X(3872)$ 
is generated dynamically by a strong coupling of the bare $\chi'_{c1}$
charmonium to the $D\bar{D}^*$ hadronic channel, with a large admixture of the
$D \bar{D}^*$ molecular component.

\newpage

\subsection{ X(3872) Decays to Quarkonia in XEFT}
\addtocontents{toc}{\hspace{2cm}{\sl T.~ Mehen}\par}

\vspace{5mm}

\noindent
Thomas Mehen

\vspace{5mm}

\noindent
Duke University, Durham, North Carolina, 27708, USA

\vspace{5mm}

There is strong  experimental evidence that the X(3872) is a very weakly bound state of $D^0 \bar{D}^{*0}+\bar{D}^0 D^{*0}$ mesons.  XEFT \cite{Fleming:2007rp}, an effective theory  of nonrelativistic $D^0$, $D^{*0}$ and $\pi^0$ mesons,  can be used to systematically calculate properties of the X(3872). Power counting shows that pion exchanges can be treated perturbatively. Leading order calculations of $X(3872) \to D^0 \bar{D}^{0}\pi^0$ in XEFT reproduce results obtained using effective range theory (ERT) \cite{Voloshin:2003nt}. XEFT can be used to systematically analyze corrections to ERT from higher dimension operators and  $\pi^0$ exchange. Effects of $\pi^0$ exchange turn out to be quite small. 

Many of the observed decays of the X(3872) are to final states with charmonium. These decays are sensitive to 
short distance aspects of the X(3872) and are therefore not completely calculable, but factorization theorems for these decays
can be developed~\cite{Braaten:B2005jj,Braaten:B2006sy}. These factorization theorems can be derived by 
matching Heavy Hadron Chiral Perturbation Theory (HH$\chi$PT) amplitudes onto XEFT operators~\cite{Fleming:2008yn}.  Decays of X(3872) to $\chi_{cJ}$ plus pions are interesting because heavy quark symmetry (HQS)  makes predictions for  the relative rates for decays to states with different $J$. If measured the relative rates could be be used to test the molecular interpretation of the X(3872)~\cite{Dubynskiy:2007tj}. In XEFT long distance contributions to these decays can significantly modify HQS predictions for the relative decay rates~\cite{Fleming:2008yn}.  Another interesting decay is the recently observed radiative transition $X(3872) \to \psi(2S) \gamma$~\cite{:2008rn}. The observed ratio for 
$\Gamma[X(3872) \to \psi(2S) \gamma]/\Gamma[X(3872) \to J/\psi \gamma]$
is a puzzle for molecular models of the X(3872). XEFT cannot address this problem because the photon in  $X(3872) \to J/\psi \gamma$ is too energetic for XEFT to be applicable. However, the polarization of $\psi(2S)$ in $X(3872) \to \psi(2S) \gamma$ is calculable. Measurement of this polarization can be used to distinguish different mechanisms that contribute to the decay \cite{ms}.

\newpage

\subsection{Scattering properties of the $X(3872)$ }
\addtocontents{toc}{\hspace{2cm}{\sl H.-W. ~Hammer}\par}

\vspace{5mm}

\noindent
Hans-Werner Hammer

\vspace{5mm}

\noindent
Helmholtz-Institut f\"ur Strahlen- und Kernphysik (Theorie)\\
and Bethe Center for Theoretical Physics,
Universit\"at Bonn, 53115 Bonn, Germany

\vspace{5mm}

Both the mass (just below the $D^{*0}\bar{D}^0$ threshold)
and the likely quantum numbers ($J^{PC}=1^{++}$) 
of the $X(3872)$ suggest that it is
either a 
weakly-bound hadronic ``molecule'' ($X(3872) \sim 1/\sqrt{2} [
D^{*0}\bar{D}^0 + \bar{D}^{*0}D^0 ]$) 
or a virtual state of charmed mesons
(See Ref.~\cite{Braaten:B2008nv} and references therein).
Assuming the $X(3872)$ is
a weakly-bound molecule, the scattering of neutral $D$ and $D^*$ mesons 
off the $X(3872)$ can be predicted from the $X(3872)$ binding energy.
We calculate the phase shifts and cross section for
scattering of $D^0$ and $D^{*0}$ mesons 
and their antiparticles off the 
$X(3872)$ in an effective field theory for short-range 
interactions \cite{Canham:2009zq}.
The total cross section is dominated by S-wave scattering of the 
$X$ and the $D^{(*)0}$ mesons. For the central value 
$E_X = 0.26 \mbox{ MeV}\,$ of the $X(3872)$ binding energy, 
the total cross section at
threshold will be of the order 800 barns for $D^0 X$ scattering
and 2600 barns for $D^{*0} X$ scattering.
This provides another example of a three-body process, along with those
in nuclear and atomic systems, that displays universal properties
\cite{Braaten:B2004rn}.
It may be possible to extract the scattering  within the 
final state interactions
of $B_c$ decays and/or other LHC events. 

This work was done in collaboration with David Canham and Roxanne Springer.
It was supported in part by the DFG through
SFB/TR 16 \lq\lq Subnuclear structure of matter,'' the BMBF
under contract No. 06BN411.

\newpage

\subsection{Prominent candidates of hidden and open charm hadronic
molecules}
\addtocontents{toc}{\hspace{2cm}{\sl F.-K. ~Guo}\par}

\vspace{5mm}

\noindent
Feng-Kun Guo

\vspace{5mm}

\noindent
Institut f\"{u}r Kernphysik
             and J\"ulich Center for Hadron Physics, Forschungszentrum J\"{u}lich,
             D--52425 J\"{u}lich, Germany

\vspace{5mm}

In this talk, I discussed how can we identify hadronic molecules and
then apply the methods to specific examples. For an $S$-wave loosely
bound state, the binding energy, and hence the scattering length,
determines the coupling constant of the bound state to its
constituents completely~\cite{molecule1,molecule2}. While the
coupling constant can be used to calculate the decay width and the
line shape of the hadronic molecule, which can be measured
experimentally, the scattering length can be simulated in lattice
QCD. This method was used to show that the present data support the
interpretation of the vector $Y(4660)$ observed in the
$\psi'\pi^+\pi^-$ mass distribution as a $\psi'f_0(980)$ bound
state~\cite{Guo:2008zg}. We also calculated the isospin-violating
decay width of the $D_{s0}^*(2317)$ in the hadronic molecular
picture~\cite{Guo:2008gp}. As a result of the large coupling of the
$D_{s0}^*(2317)$ to the constituents $DK$, the neutral and charged
meson mass differences in loops play an important role, and the
resulting value of the decay width $\Gamma(D_{s0}^*(2317)\to
D_s\pi^0)=180\pm110$~keV is much larger than that given in the
$c{\bar s}$ and tetraquark pictures. Furthermore, we show the
predicted quark mass dependence of the scattering lengths between
the charmed and light mesons agree well with the lattice
simulations~\cite{Guo:2009ct}. For heavy flavor hadronic molecules,
heavy quark spin symmetry gives us a new approach to test the
hadronic molecule assumptions of some newly observed open and hidden
charm (and also bottom in the future) resonances~\cite{Guo:2009id}.
As an application, we predicted that there should be an
$\eta_c'f_0(980)$ bound state, were the $Y(4660)$ a $\psi'f_0(980)$
bound state, with a mass of $4616^{+5}_{-6}$~MeV and the prominent
decay mode $\eta_c'\pi\pi$~\cite{Guo:2009id}. The width is predicted
to be $\Gamma(\eta_c'\pi\pi)=60\pm30$~MeV. We suggest to search it
in the $B^\pm\to K^\pm\eta_c'\pi^+\pi^-$ decays.

\newpage

\subsection{Radiative and isospin-violating decays of $D_s$-mesons  in the
hadrogenesis conjecture}
\addtocontents{toc}{\hspace{2cm}{\sl M.F.M.~ Lutz ~and~ M. ~Soyeur}\par}

\vspace{5mm}

\noindent
Matthias F.M. Lutz$\,^1$ and M. Soyeur$\,^2$

\vspace{5mm}

\noindent
$^1$Gesellschaft f\"ur Schwerionenforschung GSI,
D-64220 Darmstadt, Germany

\noindent
$^2$Irfu/SPhN, CEA/Saclay, F-91191 Gif-sur-Yvette Cedex, France

\vspace{5mm}
The hadrogenesis conjecture  was first formulated for the baryon spectrum
in \cite{ref1} and generalized to the meson spectrum in
\cite{ref2,ref3}. In both cases
the spectrum is conjectured to be the consequence of final state
interactions of preselected hadronic degrees freedom with quantum numbers
$J^P=0^-,1^-$ and
$\frac{1}{2}^-,\frac{3}{2}^-$.

In this talk we focus on the spectrum and isospin-violating strong
decays of charmed mesons with strangeness. In the heavy-light sector
of QCD besides the chiral symmetry of the light quark there are
constraints from the heavy-quark symmetry.  The latter symmetry groups
the pseudoscalar and vector D mesons with $J^P=0^-$ and $1^-$ into a
common multiplet. At leading order their masses are degenerate. Only
after the demonstration that axialvector states are generated by
chiral coupled-channel dynamics \cite{ref4}, the way was paved for an
application of the hadrogenesis conjecture to heavy-light systems
\cite{ref5,ref6}. A simultaneous study of scalar and axialvector
states is mandatory if the heavy-quark symmetry of QCD is to be kept.

The scalar D$_{s0}^*$(2317)$^\pm$ and the axial
vector D$_{s1}^*$(2460)$^\pm$ states are generated by
coupled-channel dynamics based on the leading order chiral
Lagrangian. The effect of chiral corrections is investigated. We
show that taking into account large-N$_c$ relations implies a
measurable signal for an exotic axial vector state in the $\eta D^*$
invariant mass distribution. The hadronic decay widths of the
scalar D$_{s0}^*$(2317)$^\pm$ and the axial
vector D$_{s1}^*$(2460)$^\pm$ are predicted to be 140 keV \cite{ref6}.

\newpage

\subsection{Exotic Multiplets from unitarized chiral amplitudes}
\addtocontents{toc}{\hspace{2cm}{\sl E. Oset}\par}

\vspace{5mm}

\noindent
Eulogio Oset

\vspace{5mm}

\noindent
Departamento de Fisica Teorica and IFIC, University of Valencia, Spain.

\vspace{5mm}

Using the hidden gauge formalism for pseudoscalar and vector meson interactions
\cite{Bando:1987br} a study is made of the interaction of pseudoscalar 
mesons with or without charm, leading to
scalar resonances with open and hidden charm  \cite{Gamermann:2006nm}.
Similarly, the interaction of vector mesons with pseudoscalars also leads to
dynamically generated mesons, with open and hidden charm. The X(3872) resonance
is one of those dynamically generated \cite{Gamermann:2007fi}, but we obtain two
resonances with different C-parity. 

  The interaction of vector mesons among themselves also  leads to a new sort
of states. Those with open charm can be identified with known resonances as the
$D_2^*(2460)$ and the $D^*(2640)$ \cite{Molina:2009eb}, the last one without 
experimental spin and parity assignment. In the hidden charm sector one
finds several X,Y, Z resonances around 3960-4200 MeV \cite{Molina:2009ct}.

   Finally, a thorough study of the X(3872) is made taking into account 
 $D \bar{D^*}+ cc$ components neutral and charged, together with other coupled
 channels and paying special attention to the exact masses of the particles and 
 the width of the $D^*$ \cite{Gamermann:2009fv}, making a comparison with the two
 recent experiments on $J/\psi~ \pi \pi$ and $D^0 \bar{D}^{*0}+cc$ decay from 
  Babar and Belle.

\newpage

\subsection{ Is the $X(3872)$ Production Cross Section at Tevatron Compatible with a 
 Hadron Molecule Interpretation?}
\addtocontents{toc}{\hspace{2cm}{\sl A. ~Polosa}\par}

\vspace{5mm}

\noindent
AD Polosa

\vspace{5mm}

\noindent
 INFN Roma `La Sapienza', Piazzale Aldo Moro 2, I-00185 Roma, Italy

\vspace{5mm}

The $X(3872)$ is universally accepted to be an exotic hadron. In this
letter we assume that the $X(3872)$ is a $\mol$ molecule, as claimed
by many authors, and attempt an estimate of its prompt production
cross section at Tevatron.  A comparison with CDF data allows to draw
rather compelling quantitative conclusions about this
statement~\cite{noi}. I particular we have simulated the production of
open charm mesons in high energy hadronic collisions at the
Tevatron. The generated samples have been examined searching for $D$
and $D^*$ mesons being in the conditions to form, through resonant
scattering, bound states with binding energy as small as $\sim
0.25$~MeV. These $X(3872)$ candidates have been required to pass the
same kinematical selection cuts used in the CDF data analysis. This
allows to estimate an upper bound for the theoretical prompt
production cross section of $X(3872)$ at CDF. Averaging the results
obtained with Pythia and Herwig we find this to be approximately
$0.085$~nb in the most reasonable region of center of mass relative
momenta $[0,35]$~MeV of the open charm meson pair constituting the
molecule.  This value has to be compared with the lower bound on the
experimental cross section, namely $3.1\pm 0.7$~nb, extracted from CDF
data The intuitive expectation that $S-$wave resonant scattering is
unlikely to allow the formation of a loosely bound $\mol$ molecule in
high energy hadron collision is confirmed by this analysis.

\newpage

\subsection{Impact of $D$ meson loops on charmonium decays}
\addtocontents{toc}{\hspace{2cm}{\sl Q. ~Zhao}\par}

\vspace{5mm}

\noindent
Qiang Zhao

\vspace{5mm}

\noindent
 Institute of High Energy Physics, Chinese Academy of
Sciences, Beijing 100049, P.R. China 

\noindent
 Theoretical Physics Center for Science Facilities, CAS, Beijing
100049, P.R. China

\vspace{5mm}

The obvious discrepancies between BES~\cite{bes-3770} and
CLEO-c~\cite{cleo-exclusive} on the measurement of the $\psi(3770)$
non-$D\bar{D}$ decay branching ratios bring the question of dynamics
for gluon hadronizations in $\psi(3770)\to light hadrons$.
Meanwhile, although NRQCD calculations for $\psi(3770)\to light
hadrons$ to next-to-leading order (NLO) indicate a non-negligible
branching ratio for $\psi(3770)$ non-$D\bar{D}$ decays, it may
suggest a possible failure of perturbation expansion due to large
QCD corrections from NLO~\cite{He:2008xb}. Since $\psi(3770)$ is
close to the $D\bar{D}$ open channel, we expect that the open
channel effects would be important. Thus, we propose a
non-perturbative transition mechanism via intermediate $D$ meson
loops for $\psi(3770)\to VP$~\cite{Zhang:2009kr}. By identifying the
leading meson loop transitions and constraining the model parameters
with the available experimental data for $\psi(3770)\to J/\psi\eta$,
$\phi\eta$ and $\rho\pi$, we succeed in making a quantitative
prediction for all $\psi(3770)\to VP$ with $BR_{VP}$ from $0.41\%$
to $0.64\%$. It indicates that the OZI-rule-evading long-range
interactions are playing a role in $\psi(3770)$ strong decays, and
could be a key towards a full understanding of the mysterious
$\psi(3770)$ non-$D\bar{D}$ decay mechanism.

Such a mechanism may be useful for our understanding of the
long-standing ``$\rho\pi$ puzzle'' since $\psi^\prime$ is also close
to the open $D\bar{D}$ threshold. Based on a systematic
investigation of $J/\psi(\psi^\prime)\to
VP$~\cite{Zhao:2008eg,Li:2007ky}, we identify the role played by the
short-range $c\bar{c}$ annihilation, electromagnetic (EM) transition
and intermediate meson loop transitions, which are essential
ingredients for understanding the $J/\psi$ and $\psi^\prime$
couplings to $VP$. We show that on the one hand, the EM transitions
have relatively larger interferences in $\psi^\prime\to \rho\pi$ and
$K^*\bar{K}+c.c.$ as explicitly shown by vector meson dominance
(VMD). On the other hand, the strong decay of $\psi^\prime$ receives
relatively larger destructive interferences from the intermediate
meson loop transitions. By clarifying these mechanisms in an overall
study of $J/\psi(\psi^\prime)\to VP$, we provide a coherent
prescription of the ``$\rho\pi$ puzzle''.

In brief, we present a coherent study of charmonium decays of
$J/\psi$, $\psi^\prime$ and $\psi(3770)\to VP$. It shows that the
open channel effects could be a key for understanding some of those
long-standing questions in charmonium decays. Further theoretical
studies of charmonium radiative decays~\cite{Li:2007xr}, and
isospin-violating transitions, such as $\psi^\prime\to
J/\psi\pi^0$~\cite{Guo:2009wr} and $\psi^\prime\to h_c\pi^0$, would
be useful for providing further evidence for such a mechanism.
Experimental data from BESIII in the near future would be very
helpful for justifying this idea.

\newpage

\subsection{Interplay of Quark and Meson Degrees of Freedom}
\addtocontents{toc}{\hspace{2cm}{\sl Yu. S.~ Kalashnikova}\par}

\vspace{5mm}

\noindent
Yulia S. Kalashnikova

\vspace{5mm}

\noindent
Institute of Theoretical and Experimental Physics,
117218, B.Cheremushkinskaya 25, Moscow, Russia

\vspace{5mm}

A method to identify hadronic molecules is discussed, based on 
model--independent analysis of $S$--wave low--energy hadronic 
observables, 
suggested in \cite{weinberg}, \cite{morgan}, and generalized in \cite{BHM} 
and \cite{evi}. The formalism is applied if the momenta involved are much 
smaller than the inverse range of force.
In this case the parameters entering Flatt{\`e} formula for the 
scattering amplitude contain information on the nature of the 
near--threshold resonance. In particular, it is shown that 
Weinberg--Flatt{\`e} analysis of production differential rates 
provides a direct measure for the admixture of a bare $q \bar q$    
state in the resonance wavefunction.

\newpage

\subsection{Standard charmonium vectors}
\addtocontents{toc}{\hspace{2cm}{\sl G. ~Pakhlova}\par}

\vspace{5mm}

\noindent
Galina Pakhlova

\vspace{5mm}

\noindent
ITEP, Moscow, Russia

\vspace{5mm}

The first charmonium state $J/\psi(1S)$, the bound system consisting
of the charmed quark $c$ and anti-quark $\overline c$, was discovered
in 1974. Nine more charmonium states, the $\eta_c(1S)$,
$\chi_{c0}(1P)$, $\chi_{c1}(1P)$, $\chi_{c2}(1P)$, $\psi(2S)$,
$\psi(3770)$, $\psi(4040)$, $\psi(4160)$ and $\psi(4415)$ were
observed shortly afterwards. Some of them, the so called $\psi$ states
with quantum numbers $J^{PC}=1^{--}$, were found in $e^+e^-$
annihilation. Four observed $\psi$ resonances have masses above open
charm threshold. During the next two decades no other charmonium
states were found.  A new charmonium era started in 2002. During the
past six years numerous charmoniumlike states were discovered. Among
them, only the $h_c(1P)$, $\eta_c(2S)$ and $Z(3930)\equiv
\chi_{c2}(2P)$ have been identified as candidates for conventional
charmonium, while a number of other states with masses above open
charm threshold have serious problems with a charmonium
interpretation. In particular the nature of the whole family of
charmonium-like states, found in $e^+e^-\to\pi^+\pi^- J/\psi
(\psi(2S)) \gamma_{\mathrm{ISR}}$ processes, with quantum numbers
$J^{PC}=1^{--}$ remains unclear.  Among them are the $Y(4260)$ state
observed by BaBar~\cite{babar:y4260,babar:y4260_08}, confirmed by
CLEO~\cite{cleo:y4260_isr,cleo:y4260_scan} and
Belle~\cite{belle:y4260}; the $Y(4350)$ discovered by
BaBar~\cite{babar:y4350} and confirmed by Belle~\cite{belle:y4350};
two structures, the $Y(4008)$ and the $Y(4660)$ seen by
Belle~\cite{belle:y4260,belle:y4350}.

The observation of the Y family motivated numerous measurements of
exclusive $e^+e^-$ cross sections for charmed hadron final states near
threshold.  Most of them were performed at $B$-factories using
initial-state radiation.  Belle presented the first results on the
$e^+e^-$ cross sections to the $D \overline D$, $D^+ D^{*-}$, $D^{*+}
D^{*-}$, $D^0 D^- \pi^+$ (including the first observation of
$\psi(4415) \to D \overline D{}^{*}_2(2460)$
decays)~\cite{belle:dd,belle:dst,belle:4415} and $\Lambda_c^+
\Lambda_c^-$ final states~\cite{belle:x4630}. BaBar measured $e^+e^-$
cross sections to $D \overline D$ and recently to the $D \overline
D{}^*$, $D^* \overline D{}^*$ final
states~\cite{babar:dd,babar:dd_new}.  CLEO-c performed a scan over the
energy range from 3.97 to 4.26 GeV and measured exclusive cross
sections for the $D \overline D$, $D \overline D{}^*$, and $D^*
\overline D{}^*$ final states at thirteen points with high
accuracy~\cite{cleo:cs}. The measured open charm final states nearly
saturate the total cross section for charm hadron production in
$e^+e^-$ annihilation in the $\sqrt{s}$ region up to $\sim 4.3$ GeV.

No clear evidence for open charm production associated with any of Y
states has been observed. In fact the $Y(4260)$ peak position appears
to be close to a local minimum of both the total hadronic cross
section~\cite{bes:cs} and of the exclusive cross section for $e^+e^-
\to D^* \overline D{}^*$~\cite{belle:dst,babar:dd_new}. The $X(4630)$,
recently found in the $e^+e^- \to \Lambda_c^+ \Lambda_c^-$ cross
section as a near-threshold enhancement~\cite{belle:x4630}, has a mass
and width (assuming the $X(4630)$ to be a resonance) consistent within
errors with those of the $Y(4660)$.  However, this coincidence does
not exclude other interpretations of the $X(4630)$, for example, as
the conventional charmonium state~\cite{x4630:charm,y:charm} or as a
baryon-antibaryon threshold effect~\cite{dibaryon}.

The absence of open charm decay channels for $Y$ states, large partial
widths for decay channels to charmonium plus light hadrons and the
lack of available $J^{PC}=1^{--}$ charmonium levels are inconsistent
with the interpretation of the $Y$ states as conventional charmonia.
To explain the observed peaks, some models assign the $3^3D_1(4350)$,
$5^3S_1(4660)$ with shifted masses~\cite{y:charm}, other explore
coupled-channel effects and rescattering of charm
mesons~\cite{voloshin:rescattering}.  More exotic suggestions include
hadro-charmonium~\cite{hadroch}; multiquark states, such as a
$[cq][\overline{cq}]$ tetraquark~\cite{y4260:tetra} and $D \overline
D{}_1$ or $D^0 \overline D{}^{*0}$ molecules~\cite{y4260:molecule}.
One of the most popular exotic options for the $Y$ states are the
hybrids expected by LQCD in the mass range from $4.2-5.0$
GeV~\cite{y4260:hybryds}.  In this context, some authors expect the
dominant decay channels of the Y(4260) to be $ Y(4260)\to
D^{(*)}\overline D{}^{(*)}\pi$.

Recently Belle reported the first measurement of the $e^+e^- \to D^0
D^{*-} \pi^+ $ exclusive cross section at
threshold~\cite{belle:ddstpi}.  The values of the amplitude of the
$Y(4260)$, $Y(4350)$, $Y(4660)$ and $X(4630)$ signal function obtained
in the fit to the $M_{D^0D^{*-} \pi^+}$ spectrum are found to be
consistent with zero within errors.  Belle found no evidence for
$Y(4260)\to D^0 D^{*-} \pi^+ $ decays as predicted by hybrid models
and obtained the upper limit on $Br(Y(4260) \to D^0 D^{*-} \pi^+)/Br(
Y(4260) \to \pi^+ \pi^- J/\psi) < 9$ at the 90\% C.L.

\newpage

\subsection{Double charmonium production in $\mathbf {e^+e^-}$, new
  states and unexpectedly large cross sections}
\addtocontents{toc}{\hspace{2cm}{\sl P. ~Pakhlov}\par}

\vspace{5mm}

\noindent
Pasha Pakhlov

\vspace{5mm}

\noindent
Institute for Theoretical and Experimental Physics,
Moscow, Russia

\vspace{5mm}

Prompt charmonium production in $\ee$ annihilation is important for
studying the interplay between perturbative QCD and non-perturbative
effects. The production rate and kinematic characteristics of
\jp\ mesons in $\ee$\ annihilation are poorly described by theory, and
even the production mechanisms are not well understood. An effective
field theory, non-relativistic QCD (NRQCD), based on leading order
perturbative QCD calculations, predicted that prompt \jp\ production
at $\sqrt{s}\!\approx \! 10.6$\gev\ is dominated by \eejpgg\ with a
$1\,$pb cross section~\cite{nrqcd}; the \eejpg\ contribution may be of
the same order, is uncertain due to poorly-constrained color-octet
matrix elements~\cite{nrqcd1}. The \eejpcc\ cross section is predicted
to be $\sim \! 0.05-0.1\,$pb~\cite{Kiselev:94}.

$\gtrsim$
By contrast, in 2002 Belle observed the ratio of the \jpcc\ and
inclusive \jp\ production cross sections to be $0.59^{+0.15}_{-0.13}
\pm 0.12$~\cite{Belle:2cc}, and thus found
$\sigma$(\eejpcc)$\gtrsim \sigma$(\eejpgg). In 2009, using an order of
magnitude larger data sample Belle measured the cross sections for the
processes \eejpcc\ and \jpncc\ in a model independent
way~\cite{Belle:cc2}. The measured cross sections are $(0.74 \pm
0.08{\,}^{+0.09}_{-0.08})$\,pb and $(0.43 \pm 0.09 \pm 0.09)$\,pb,
respectively, thus the last measurements confirmed that \eejpcc\ is
the dominant mechanism for \jp\ production in $\ee$ annihilation,
contrary to earlier NRQCD predictions. Recently, both \eejpgg\ and
\jpcc\ cross sections have been recalculated including NLO corrections
and are in better agreement with the experimental data\,\cite{Ma:09,
  Gong:09}. However, the measured \eejpcc\ cross section exceeds the
perturbative QCD prediction $\sigma(e^+e^- \! \to \!
c\bar{c}c\bar{c}) \!  \approx \!\!  0.3$\,pb~\cite{Berezhnoy:07},
which includes the case of fragmentation into four charmed hadrons,
rather than \jp \cc.

The \eejpcc\ process is dominated by \cc\ fragmentation to open charm,
with a $(16 \pm 3)\%$ contribution from double charmonium production,
{\emph i.e.} production of a second charmonium below the open charm
threshold in the event with \jp. The large rate for processes of the
type \eejpet\ reported by Belle ($\sigma($\eejpet$)= (25.6\pm 2.8\pm
3.4)\,$fb)~\cite{Belle:2cc, Belle:2cc2}, also remained a puzzle for
many years. The first NRQCD calculations~\cite{Braaten:B03} gave at
least an order of magnitude smaller value ($\sim \!2$\,fb) than those
measured by Belle. The importance of relativistic corrections was
realized in Ref.~\cite{Ma:04, Bondar:05}; the authors, using light
cone approximation to take into account the relative momentum of heavy
quarks in the charmonium, managed to calculate the cross section which
is close to the experimental value. Alternatively, authors of
Ref.~\cite{Bodwin:08} suggested to resolve the discrepancy within the
NRQCD approach by the resummation of relativistic corrections,
contribution from pure QED diagram, the corrections of next-to-leading
order in $\alpha_s$.

Double charmonium production in$\ee$ annihilation can be used to
search for new charmonium states with charge conjugation $C\!=\!+1$,
recoiling against known and easily reconstructed $C\!=\!-1$ charmonium
mesons such as the \jp\ or \pp. Studies of various double charmonium
final states have demonstrated that scalar and pseudoscalar charmonia
are produced copiously recoiling against a \jp\ or \pp\ and there is
no significant suppression of the production of radially excited
states. In 2008, Belle observed the processes \eedd\ (\dds, \dsds) and
reported the observation of the clear enhancement with a significance
of $5.1\,\sigma$ in the invariant mass distribution of
\dsds\ combinations in the process \eedsds, which was interpreted as a
new charmonium-like state, the $X(4160)$~\cite{Belle:x4160}. The
$X(4160)$ parameters are $M= (4156\,^{+\,25}_{-\,20} \pm
15)$\,\mevc\ and $\Gamma = (139\,^{+\,111}_{-\,\phantom{1}61} \pm
21)$\,\mev. Belle also confirmed the observation of the
charmonium-like state, $X(3940)\!  \to \! D \overline{D}{}^*$,
produced in the process \eejpxn\ with a significance of
$5.7\,\sigma$. The $X(3940)$ mass and width are
$M=(3942\,^{+\,7}_{-\,6} \pm 6)$\,\mevc and
$\Gamma=(37\,^{+\,26}_{-\,15}\pm 8)$\,\mev, consistent with the first
Belle result~\cite{Belle:x3940}.

If the $X(3940)$ has $J=0$ the absence of a \dd\ decay mode strongly
favors $J^{P}=0^{-+}$, for which the most likely charmonium assignment
is the $\eta^{\prime\prime}_c$. The fact that the lower mass $\eta_c$
and $\eta_c^{\prime}$ are also produced in double charm production
supports this assignment. However, there is a problem that the
measured mass of the $X(3940)$ is below potential model estimates of
$\sim\!4050$\mevc\ or higher~\cite{Barnes:05}. A further complication
is the observation by Belle of $X(4160)$, which could also be
attributed to the $3^1S_0$ state, using similar arguments. But the
$X(4160)$ mass is well above expectations for the $3^1S_0$ and well
below those for the $4^1S_0$, which is predicted to be near
4400\,\mevc~\cite{Barnes:05}. Although either the $X(3940)$ or the
$X(4160)$ might conceivably fit a charmonium assignment, it seems very
unlikely that both of them could be accommodated as \cc\ states.

\newpage

\subsection{Meson spectroscopy via ISR}
\addtocontents{toc}{\hspace{2cm}{\sl S. ~Pacetti}\par}

\vspace{5mm}

\noindent
Simone Pacetti

\vspace{5mm}

\noindent
Enrico Fermi Center, Rome, Italy\\ 
INFN, Laboratori Nazionali di Frascati, Frascati, Italy\\

\vspace{5mm}
The initial state radiation technique (ISR) allows to exploit flavor factories
as usual $\ee$\ experiments with energy scan in the center of mass. 
A generic hadronic state $X_{\rm had}$, with mass lower than the fixed energy of the 
machine, can be studied through the process $\ee\to X_{\rm had}\gamma_{IS}$, 
where the photon $\gamma_{IS}$ is emitted by one the initial leptons. In
Born approximation $X_{\rm had}$ has quantum numbers $J^{PC}=1^{--}$. Using the
ISR technique \bbr\ discovered the first element of the charmonium-like $Y$ family, 
i.e. the \yy~\cite{b4260} in the channel $\ee\to\jpsi\pi^+\pi^- \gamma_{IS}$.
This observation was confirmed by Belle~\cite{be4260}, in the same $\jpsi\pi^+\pi^-$  
dominant decay channel, and by CLEO~\cite{cleo}, not only in $\jpsi\pi^+\pi^-$  but 
also in $\jpsi2\pi^0$ and $\jpsi K^+K^-$.
Looking for the \yy\ Belle identified also another structure with lower mass in the 
same $\jpsi\pi^+\pi^-$ channel, this additional state has been called \y~\cite{be4260}.
In 2007 \bbr\ observed another $Y$ state in the channel $\psii\pi^+\pi^-$ with
 $\psii\to\jpsi\pi^+\pi^-$. This structure, called \yyy~\cite{b4325}, emerged very 
close to the $\psii\pi^+\pi^-$ threshold, has been confirmed by Belle, in the
same channel, but with a slightly different mass~\cite{be4360}. In the same 
investigation, thanks to a larger data sample, Belle discovered also another
resonance decaying in $\psii\pi^+\pi^-$, the \yyyy~\cite{be4360}.
\\
The simplest interpretation of these resonances as the still missing
states in the charmonium spectrum of the time-honored quark model~\cite{isgur} 
has been largely disfavored by their non-observation~\cite{dd} in the charmonia
dominant decay channels, i.e. decays in charmed hadrons. Nevertheless, the $\ee\to\lclc$ 
cross section, recently measured by Belle~\cite{lclc}, shows a clear near-threshold 
enhancement that has been identified as a further $J^{PC}=1^{--}$ resonance,
the \x, which is compatible with the \yyyy.
\\
To summarize, two main classes of $Y$ resonances have been identified: the lightest
\y\ and \yy, which decay in $\jpsi\pi\pi$, and the heaviest \yyy\ and \yyyy\ decaying
only in $\psii\pi\pi$. No signals for those structures have been found in open
charm final states. The \x\ seems to escape this classification.
\\
The possible interpretation, besides the disproved charmonium states, is
based on three main ideas: hybrid charmonia~\cite{hyb}, molecules and 
hadro-charmonia~\cite{meiss}, and threshold effects~\cite{threshold}.
The relatively small widths contrast the tetraquark hypothesis.
\\
However, all the attempted interpretations have the
negative feature to consider only one $Y$ state individually.
It is, indeed, manifest that all these $J^{PC}=1^{--}$ states show 
incredibly similar properties: they share the same decay channels,
they have similar total widths, as well as similar $\Psi(1S,2S)\pi\pi$ 
branching fractions. It follows that a more comprehensive description
is needed. An attempt in this direction has been made in Ref.~\cite{screen},
where it is shown how the spectrum of higher charmonium, obtained using a
screened $c\bar{c}$-potential, describe quite well all the new $Y$ states.
%
%

\newpage

\subsection{Perspectives for spectroscopy at super-B factories}
\addtocontents{toc}{\hspace{2cm}{\sl C. ~Patrignani}\par}

\vspace{5mm}

\noindent
Claudia Patrignani

\vspace{5mm}

\noindent
Universit\`a\ di Genova and INFN Genova

\vspace{5mm}

The B-factories observed a number of charmonium-like states which do
not easily fit into the conventional charmonium picture~\cite{reviews},
The present knowledge of the properties of these states is generally
based on small samples of events. 

The large samples that would be collected
at either of the proposed super-B factories ~\cite{Hashimoto:2004sm},
\cite{Bona:2007qt}, \cite{Hitlin:2008gf} would allow to understand the nature
of these states, discriminating among the many different proposed interpretations.

\newpage

\section{Short summary of the posters}

\subsection{Franck-Condon principle for Heavy Quarkonium Decays and Heavy Quark Effective Theory}
\addtocontents{toc}{\hspace{2cm}{\sl F. J.~ Llanes-Estrada}\par}

\vspace{5mm}

\noindent
Felipe J. Llanes-Estrada 

\vspace{5mm}

\noindent
Universidad Complutense de Madrid, Departamento de F\'{\i}sica Te\'orica I.

\vspace{5mm}

 In~\cite{LlanesEstrada:2008nw} we
have proposed to adapt the Franck-Condon principle of molecular
physics to ascertain the nature of Heavy Quarkonium above open flavor
threshold in terms of its heavy quark constituents.  Our current
formulation presented at Charm-ex applies to Heavy Quarkonium decaying
to heavy mesons carrying one heavy quark each plus any number of pions
or perhaps other light degrees of freedom. The principle states that
"The velocity distribution of heavy mesons carrying one heavy quark
each and following heavy quarkonium decay, coincides with the velocity
distribution of those heavy quarks inside the heavy quarkonium".  Once
such velocity distributions have been measured experimentally, they
can provide invaluable insight into the structure of excited heavy
quarkonium, whether to study valence or sea degrees of freedom. We
have discussed possible applications with conference attendees.  This
principle we now understand to be a consequence of essentially all
heavy quark effective theories proposed~\cite{Brambilla:2004jw} to
date, such as NRQCD or HQQET. Their Lagrangian densities describe the
motion of heavy quarks whose velocity is not changed by QCD
interactions in leading order of $\Lambda_{\rm QCD}/M_Q$. We are
trying to develop a way to estimate corrections at first order (work
in progress in collaboration with Juan Torres Rinc\'on and Ignazio
Scimemi).

\newpage

\subsection{Mixing of S-wave charmonia with D\={D} molecule states}
\addtocontents{toc}{\hspace{2cm}{\sl G. ~Bali, ~C. ~Ehmann}\par}

\vspace{5mm}

\noindent
Gunnar Bali, Christian Ehmann

\vspace{5mm}

\noindent
University of Regensburg

\vspace{5mm}
One possible decay channel of charmonia is into $D\bar{D}$ molecule states by creation of a light quark-/antiquark pair. The investigation of such decays sheds light on the higher fock state contributions to the charmonia wavefunction~\cite{ref1b} and potential mass shifts.
A variational approach is applied to a mixing matrix containig both charmonia and $D\bar{D}$ molecule interpolating fields. The calculation of several diagramms appearing in this matrix requires all-to-all propagators, which are realised by sophisticated stochastic estimator techniques~\cite{ref2b}.
The runs are performed on $N_f=2$ $24^3\times 48$ with $M_{\pi} \approx 380$ MeV configurations using the non-perturbatively improved Clover-Wilson action, both for valence and sea quarks.

\newpage

\subsection{Spectra of low and high spin mesons with light quarks from lattice QCD}
\addtocontents{toc}{\hspace{2cm}{\sl T.~Burch ~et ~al.}\par}

\vspace{5mm}

\noindent
T.~Burch, C.~Hagen$\,^*$, M.~Hetzenegger, A.~Sch\"afer

\vspace{5mm}

\noindent
Universit\"at Regensburg

\vspace{5mm}

We present results for excited meson spectra from $N_f=2$ clover-Wilson configurations 
provided by the CP-PACS Collaboration. In our study we investigate both low and high spin mesons. 

For spin-0 and spin-1 mesons, we are especially interested in the excited states. To 
access these states we construct several different interpolators from quark sources of 
different spatial smearings including ones which resemble p-waves and then calculate a 
matrix of correlators. We then apply the variational method \cite{varmethod} and solve 
a generalized eigenvalue problem for this matrix.
For spin-2 and spin-3, we extract only the lowest lying states using interpolators which 
have been successful in spectroscopy calculations of charmonia \cite{Liao:2002rj}.
We are able to successfully isolate excited states in the pseudoscalar and vector channel 
and obtain a number of high spin mesons up to $J=3$.

First results have been published in Refs. \cite{firstresults}. Our final results can be found in Ref. \cite{Burch:2009wu}.

\newpage

\subsection{Charmonium Hybrids at $\overline{\mathrm{P}}$ANDA}
\addtocontents{toc}{\hspace{2cm}{\sl J. ~Schulze ~and~ M. ~Peliz\"aus}\par}

\vspace{5mm}

\noindent
J. Schulze and M. Peliz\"aus for the $\overline{\mathrm{P}}$ANDA Collaboration

\vspace{5mm}

\noindent
Ruhr-Universit\"at Bochum, Institut f\"ur Experimentalphysik I

\vspace{5mm}

A c$\overline{\mathrm{c}}$-pair bound by an excited gluonic flux tube is called a charmonium hybrid. The spin contribution of the flux tube can lead to spin-exotic states, of which the 1$^{-+}$-state (labelled $\tilde{\eta}_{\mathrm{c1}}$ in this work) is commonly expected to be the lightest with a mass between 4.1 and 4.4 GeV/c$^2$ \cite{Bernard:1997},\cite{Close:1998},\cite{Page:1998}. Open and hidden charm decays are predicted such as $\tilde{\eta}_{\mathrm{c1}}\rightarrow\chi_{\mathrm{c1}}\pi^0\pi^0$ and $\tilde{\eta}_{\mathrm{c1}}\rightarrow D\overline{D}^*$.\par
Hybrid states are believed to appear in gluon-rich processes as seen in $\overline{p}p$ annihilations. One of the focuses of the $\overline{\mathrm{P}}$ANDA experiment, which is going to be built at the antiproton storage ring (HESR) in course of the FAIR project at the GSI in Darmstadt, Germany, will be the search for gluonic excited charmonium states.\par 
To demonstrate the sensibility of the $\overline{\mathrm{P}}$ANDA experiment on detection of these states, Monte Carlo studies have been performed for $\overline{p}p\rightarrow\tilde{\eta}_{\mathrm{c1}}\eta$ production at the highest available energy $\sqrt{s}=5.47\,\mathrm{GeV}$ \cite{Lutz:2009ff}. Final states with a high photon multiplicity need to be reconstructed, wherefore an outstanding electromagnetic calorimeter is needed. In addition, a good 
identification of electrons, pions and kaons is mandatory.\par 

Possible background channels are events with a similar signature as the signal reactions (such as $J/\psi\pi^0\pi^0\pi^0\eta$ or $D^0\overline{D}^{*0}\pi^0$). Considering these reactions, a signal-to-background ratio of better than 200 ($\overline{p}p\rightarrow \tilde{\eta}_{\mathrm{c1}}\eta\rightarrow \chi_{\mathrm{c1}}\pi^0\pi^0$) and better than 4000 ($\overline{p}p\rightarrow \tilde{\eta}_{\mathrm{c1}}\eta\rightarrow D\overline{D}^*$) can be achieved. Assuming a cross section of 30 pb (calculated from the reverse reaction $\Psi(\mathrm{2S}) \rightarrow \eta \overline{\it p}\it p$ for conventional charmonium \cite{Lundborg:2005am}), the number of reconstructed events per month are 
$N=B(\tilde{\eta}_{c1}\to\chi_{c1}\pi^0\pi^0)\times 5$ and $N=B(\tilde{\eta}_{c1}\to D^0\overline{D}^{*0})\times 2$.

Thus, to detect the investigated reactions having low cross sections, the $\overline{\mathrm{P}}$ANDA experiment in conjunction with the high luminosity of HESR is well suited.

\begin{flushleft}

\end{flushleft}

\newpage

\subsection{The \bes~Tau-Charm Factory}
\addtocontents{toc}{\hspace{2cm}{\sl M.~Pelizaeus~ and ~J.~Zhong}\par}

\vspace{5mm}

\noindent
M.~Pelizaeus and J.~Zhong (for the \bes~Collaboration)

\vspace{5mm}

\noindent
Ruhr-Universit\"at Bochum, Institut f\"ur Experimentalphysik I

\vspace{5mm}
The \bes~experiment \cite{Asner:2008nq} at the symmetric electron
positron storage ring \bepcii~(Beijing) running in the energy range
$\sqrt{s}=2\ldots4.4\;\mathrm{GeV}$ has started its operation in summer
2008. The high luminosity of the machine in conjunction with the good
tracking, particle identification and calorimetry of the detector
offers excellent opportunities for light and charmed hadron
spectroscopy, the study of charmonium transitions, $D\bar{D}$ mixing,
$CP$-violation in $D$ meson decays, $\tau$ physics and other topics.
During the first run periods in March-April and June-July 2009 data
samples corresponding to more than $1\cdot10^{8}$ $\psi(2S)$ and
$2\cdot10^{8}$ $J/\psi$ events, respectively, have been recorded.

Preliminary results on fully reconstructed
$\psi(2S)\to\chi_{cJ}\gamma$ decays, with $\chi_{cJ}$ decaying into
$\pi^+\pi^-\pi^+\pi^-$, $K^+K^-K^+K^-$, and $\pi^+\pi^-p\bar{p}$ have
been presented. Clean $\chi_{cJ}$ signals for all three hadronic final
states are observed, demonstrating the capabilities of the
detector. Also an inclusive study of $\psi(2S)\to h_c\pi^0$,
$h_c\to\gamma\eta_c$ decays has been performed. Events are tagged by
the radiative photon. The $h_c$ is identified in the
recoil spectrum of the reconstructed $\pi^0$, confirming the
observation of the $h_c$ recently reported by the CLEO collaboration
\cite{Rosner:2005ry,Rubin:2005px}.

\newpage

\subsection{The $\overline{\mathrm{P}}$ANDA Electromagnetic Calorimeter}
\addtocontents{toc}{\hspace{2cm}{\sl J. ~Becker ~and ~T.~ Held}\par}

\vspace{5mm}

\noindent
J\"orn Becker und Thomas Held

\vspace{5mm}

\noindent
Ruhr-Universit\"at Bochum, Institut f\"ur Experimentalphysik I

\vspace{5mm}

Antiproton-proton annihilations in the high energy antiproton storage ring (HESR) of the future FAIR facility in Darmstadt will allow sensitive tests of quantum chromo dynamics. The $\overline{\mathrm{P}}$ANDA detector aims at precision studies of charm-quark mesons, glue balls and hybrid mesons. One of the crucial detector components for those studies is the electromagnetic calorimeter \cite{PANDA PPR}. Its overall concept is presented in the Technical Design Report for the $\overline{\mathrm{P}}$ANDA EMC \cite{PANDA EMC TDR}. The $\overline{\mathrm{P}}$ANDA calorimeter consists of about 15000 lead tungstate crystals in a barrel part, one forward and one backward endcap covering - in combination with a forward spectrometer - 99\% of the whole solid angle. The scintillator material is radiation hard, allows a compact design, a high rate capability, a low energy threshold of 10 MeV, and an energy range up to 15 GeV. The energy resolution of the calorimeter is expected to be \[\frac{\sigma_E}{E}\leq 1\%\oplus \frac{2\%}{\sqrt{E/GeV}}\]. 

In order to raise the light yield of the scintillator material the operation temperature of the calorimeter will be -25$^o$C. The result is a factor of four more light compared to an operation at room temperature. Unfortunately the temperature dependence of the light yield at -25$^o$C increases to $dLY/dT=3\%$. Therefore temperature stability is essential for a high energy resolution. A sophisticated high insulating airtight casing is needed to keep the temperature stable and to prevent icing on the crystals. The thermal requirements are a temprature variation of less than 0.1$^o$C and a temperature inhomogeneity along a crystal of less than 0.1$^o$C per centimeter.

The monitoring of temperature and humidity inside the calorimeter is done by 'Temperature and Humidiy monitoring boards for $\overline{\mathrm{P}}$ANDA' (THMPs), designed for the readout of 64 channels each and rated for radiation doses up to 700 Gy at an operating range of -30$^o$C to +30$^o$C. PT 100 platinum temperature sensors with a thickness of only 60 $\mu m$, fitting in the space between adjacend crystals, have been developed. The achievable precision of the temperature monitoring is 0.05$^o$C.
 
There is a 192 crystal calorimeter endcap prototype under construction that will allow tests of the components and its composition under $\overline{\mathrm{P}}$ANDA operating conditions.

\begin{flushleft}

\end{flushleft}

\newpage

\section{List of Participants}

\begin{itemize}
\item Baru Vadim, 	J\"ulich Center for Hadron Physics			
\item Becker	J\"orn,		Ruhr-Universit\"at Bochum
\item Cleven	Martin,		J\"ulich Center for Hadron Physics
\item Denig	Achim,	Universit\"at Mainz
\item Dudek	Jozef,	Old Dominion Universtiy/J-Lab
\item Ehmann	Christian,		Universit\"at Regensburg
\item Eidelmann	Simon,	Budker Institute of Nuclear Physics, Novosibirsk
\item Fritsch	Miriam,		Universit\"at Mainz
\item Gillitzer	Albrecht,		J\"ulich Center for Hadron Physics
\item Guo	Feng-Kun,	J\"ulich Center for Hadron Physics
\item Hagen	Christian,		Universit\"at Regensburg
\item Hammer	Hans-Werner,		Universit\"at Bonn
\item Hanhart	Christoph,		J\"ulich Center for Hadron Physics
\item Held	Thomas,		Ruhr-Universit\"at Bochum
\item Kalashnikova	Yulia,		ITEP, Moscow
\item Krewald	Siegfried,		J\"ulich Center for Hadron Physics
\item Llanes-Estrada	Felipe,		University Computense Madrid
\item Lutz	Matthias F.M.,		GSI, Darmstadt
\item McNeile	Craig,		University of Glasgow
\item Mehen	Thomas,		Duke University
\item Metsch	Bernard,		Universit\"at Bonn
\item Mei\ss ner	Ulf-G.,		Universit\"at Bonn and J\"ulich Center for Hadron Physics
\item Meyer	Curtis,		Carnegie Mellon University, Pittsburgh
\item Nefediev	Alexey,		ITEP, Moscow
\item Oset	Eulogio,		University of Valencia
\item Olsen	Stephen, L.,		Seoul National University
\item Pacetti 	Simone,		Enrico Fermi Center Rome and INFN-LNF Frascati
\item Pakhlov	Pavel,		ITEP, Moscow
\item Pakhlova	Glina,		ITEP, Moscow
\item Patrignani	Claudia,		Universita' di Genova and INFN
\item Peliz\"aus	Marc,		Ruhr-Universit\"at Bochum
\item Peters	Klaus,		GSI, Darmstadt
\item Polosa	Antonio,		INFN Frascati
\item Ritman	Jim,		J\"ulich Center for Hadron Physics
\item Schulze	Jan,		Ruhr-Universit\"at Bochum
\item Solodov	Evgeny,		Budker Institute of Nuclear Physics, Novosibirsk
\item Stockmanns	Tobias,		J\"ulich Center for Hadron Physics
\item Swanson 	Eric,		University of Pittsburgh
\item Wiedner	Ulrich,		Ruhr-Universit\"at Bochum
\item Ziegler	Veronique,		SLAC National Accelerator Laboratory, Menlo Park
\item Yuan	Changzheng,		IHEP, Beijing
\item Zhao	Qiang,		IHEP, Beijing
\item Zhong	Jan,		Ruhr-Universit\"at Bochum
\end{itemize}

\end{document}